\documentstyle[twocolumn,prc,aps,epsfig]{revtex}

     \newcommand{\pathnow}{}
\begin{document}\hbadness=10000
\topmargin=-.7cm\oddsidemargin = -0.7cm\evensidemargin = -1.7cm
  \hyphenation{strang-en-ess}
\twocolumn[\hsize\textwidth\columnwidth\hsize\csname %
@twocolumnfalse\endcsname
\title{Strange hadrons and their resonances:\\  
a diagnostic tool  of  QGP  freeze-out dynamics
}
\author{Johann Rafelski$^1$, Jean Letessier$^2$,  and Giorgio Torrieri$^1$}
\address{
$^1$Department of Physics, University of Arizona, Tucson, AZ 85721\\
$^2$LPTHE, Universit\'e Paris 7
}
\date{April 11, 2001}
\maketitle
\begin{abstract}
We update our  chemical analysis of (strange) hadrons 
produced at the SPS in Pb--Pb collisions at $158A$ GeV   
and discuss chemical analysis of RHIC results. 
We report that the shape of (anti)hyperon $m_\bot$-spectra 
in a  thermal freeze-out analysis leads to freeze-out 
conditions found in chemical analysis, implying sudden
strange hyperon production. We discuss  how a combined 
analysis of several strange hadron resonances of differing 
lifespan can be used  to understand the dynamical
process present during  chemical and thermal freeze-outs. 
In medium resonance quenching is considered. \vskip 0.1cm

PACS: 12.38.Mh, 12.40.Ee, 25.75.-q
\end{abstract}
\pacs{PACS: 12.38.Mh, 12.40.Ee, 25.75.-q}
\vspace{-0.1cm}
]
\begin{narrowtext}
\section{Introduction}
The quark-gluon plasma (QGP) as we today  call hot quark matter
has been predicted many years ago to be a possible new state of 
hadronic matter.
As the ideas about QGP formation in high energy nuclear 
collisions matured, a challenge  emerged how
the locally deconfined state which exists  10$^{-22}$s
can be distinguished from the gas of confined hadrons. This is 
also a matter of principle, since arguments were advanced 
that this may be impossible. A quark-gluon
 based description could be just a change of Hilbert space 
expansion basis. However, it is believed that a change in the 
structure of matter occurs at high temperature and QGP is
qualitatively different compared to matter made of 
confined hadrons \cite{DiG00}.

Clearly, these difficult questions can be settled by an 
experiment, if a probe of QGP operational on the collision
time scale, can be devised.  Several QGP  observables were 
proposed and we  address here our recent progress in 
the development  of strangeness and strange antibaryon 
production as an observables of QGP.
Strangeness signature of QGP originates in the observation that 
when color bonds are broken, the chemically (abundance) equilibrated 
deconfined state has an unusually high abundance of strange
quarks \cite{abundance}. 

There was possibility that the  relatively small
size of the plasma fireball would suppress strangeness yield.
It was shown that when the system size
is greater than about five elementary hadronic volumes \cite{RD80}
the physical properties of the hadronic system, including 
in particular strangeness abundance, are nearly 
as expected for an infinite system. Subsequently, 
kinetic study of the dynamical process 
of chemical equilibration  has shown that the  
gluon component in the QGP is 
able to produce strangeness rapidly \cite{RM82}, allowing formation
of  (nearly) chemically equilibrated dense phase of deconfined,
hot, strangeness-rich, quark matter in relativistic nuclear collisions. 
Therefore abundant strangeness production is today generally viewed 
to be  related directly to  presence of gluons in QGP.

The high density of strangeness  in the reaction fireball 
favors formation of multi strange hadrons \cite{Raf82}, which 
are produced  rarely if only individual hadrons 
collide \cite{Koc85,KMR86}. The predicted systematics
of strange antibaryon production has
in fact been observed, rising with strangeness content \cite{WA97p}.
Moreover there is now evidence that  $\overline\Xi$ production
shows  a sudden onset when the number of participating (wounded)
nucleons exceeds 50 \cite{NA57Mor}. Similar results were reported for the
Kaon yields by the NA52 collaboration \cite{Kab99}. This threshold 
behavior arises for volumes which are large compared to the threshold
found in Ref.\cite{RD80}, thus  if the experimental results 
are trustworthy (and we have no reason to doubt them)
they show that this effect is probably not
result of the smallness of the physical system 
(`canonical suppression' see  Ref.\,\cite{Red01})
but is more likely result of opening up of novel reaction mechanisms, 
as is expected should QGP formation occur. 

Definition of the baseline when determining yield
enhancement is important. Indeed one observes for some strange 
particles  already an enhancement  comparing $pp$  (proton-proton)
 to $pA$ (proton-nucleus)   interactions. There are several natural
reasons to expect a change in  production pattern when 
comparing $pA$ with $pp$ reactions, such as isospin 
selection rules, or (anti)shadowing of participating nucleons. 
This leads to `enhancements' in non-strange particles along with 
strange particles \cite{Saf01}. 
For this reason the baseline for comparison of 
$AA$ (nucleus-nucleus) results should be always the 
$NA$ (nucleon-nucleus) collision system,
and one should show that the value of $A$ does not matter  
in $NA$ reactions which establish the baseline (i.e there is 
scaling of the yields with A). 

We also see in the experimental data we address here 
that particles of very different  properties are produced 
by the same mechanism since they are appearing with 
identical or similar $m_\bot$-spectra \cite{Ant00}.  
The  symmetry between strange
baryon and antibaryon spectra is strongly suggesting that 
the same reaction mechanism produces $\Lambda$ and $\overline\Lambda$ 
and $\Xi$ and $\overline\Xi$. This is understood readily if
 a dense fireball of deconfined matter formed in heavy ion
reactions  expands explosively, super cools, 
and in the end encounters a mechanical instability which facilitates 
sudden break up into hadrons  \cite{Raf00}. 

Another evidence for this sudden reaction mechanism arises if there is 
coincidence of chemical(particle yield)  and thermal (spectral 
shape) freeze-out. By definition at thermal freeze-out condition   
the momentum distributions of final state particles 
stop evolving after expansion dilution of dense matter  fireball 
reduces frequency of  elastic and inelastic collisions. 
Inelastic reactions occur more rarely, and they change 
hadron abundances. Thus in general chemical freeze-out naturally
occurs earlier than the thermal freeze-out. 
Simultaneous chemical and thermal freeze-out require non-equilibrium
evolution of the fireball as has been discussed recently \cite{Raf00}.

We study the chemical freeze-out conditions reached  at 
highest  SPS energy in subsection  \ref{freezechem}. 
In subsection  \ref{freezeRHIC} we show that the first RHIC 
run data does not allow  to determine 
the temperature of particle production. Turning back to SPS results, 
we  discuss in subsection \ref{freezespectra}
the results of  thermal freeze-out analysis \cite{Tor00}.  These 
results show that for  the 4 collision centralities of 
the experiment WA97 with participant number greater than 100, 
the thermal and chemical freeze-out conditions (described in
subsection \ref{freezechem})  are practically the same. 
This offers an excellent confirmation of the sudden QGP
breakup hypothesis \cite{Raf00}.

In section \ref{resonance}, we consider the production of strange 
hadron resonances as  a method to study the dynamics of QGP 
hadronization. The idea is to use
abundance of unstable resonances which have varying width and to 
determine  fraction  which becomes unobservable in
consideration of the re-scattering effects: once resonance products 
rescatter one cannot `see' the resonance by reconstruction \cite{Tor01}. 
We expand this idea here allowing for the  phenomenon that in dense matter
a resonance which is `unnaturally' narrow could be `quenched' in collisional
processes and decay much faster, which would give a greater opportunity 
for the rescattering to occur.

\section{Chemical and thermal freeze-out}\label{freeze}
\subsection{Global chemical freeze-out condition at SPS}\label{freezechem}
After a recent update of some experimental results \cite{S2000}, we have
updated our earlier chemical analysis \cite{Let00}. 
Our strategy is to maximize the precision of the 
description of the final multi-particle hadron state employing
statistical methods.   In our present
chemical freeze-out analysis there are a few theoretical refinements 
such as use of Fermi-Bose statistics throughout, more extensive 
resonance cascading. In the experimental input data compared to earlier
work we omit the   NA49 $\overline\Lambda/\bar p$ 
ratio and update the  NA49 $\phi$-yields. The total $\chi^2_{\rm T}$,
the number of measurements used $N$ the number of parameters being varied $p$ 
and the number of restrictions on data points $r$ are shown in 
heading of the  table \ref{newfitqpbs}. The values imply that our model 
of the hadronic phase space has a  very high confidence level. 

In the upper section  of table \ref{newfitqpbs}, we show statistical 
model parameters which best describe the experimental results for
Pb--Pb data. We show in turn chemical freeze-out temperature,
$T$ [MeV], expansion velocity $v$, the light and strange quark fugacities
$\lambda_{q},\lambda_{s}$ and light quark phase space 
occupancy $\gamma_{q}$ and the
strange to light quark ratio $\gamma_{s}/\gamma_{q}$. 
We fix $\gamma_{q}$ at the point of maximum pion entropy density
${\gamma_{q}^c}=e^{m_\pi/2T_f}$ \cite{Let00},
since this is the natural value to which the fit
converges once the Bose distribution for pions is used. 

It is interesting that in the Pb--Pb collisions  $\gamma_{s}/\gamma_{q}$
 is so close to unity, the often tacitly assumed value. 
In this detail the revised analysis differs more than 2 s.d. 
from our earlier results \cite{Let00}.
The only other notable new finding  is the prediction for
$\overline\Lambda/\bar p\simeq 0.6$ (not shown in table). 

In the bottom section  of table \ref{newfitqpbs}, we show 
physical properties of the fireball derived from
the properties of the hadronic phase space:  $E_{f}^{in}/S_{f}$, the 
specific energy per entropy of the hadronizing volume element 
in local rest frame;  ${s}_{f}/b$  specific strangeness per baryon;
$({\bar s}_f-s_f)/b$ net strangeness
 of  the full hadron phase space characterized by these
statistical parameters. The relevance of this results is that
 $E_{f}^{in}/S_{f}$ characterizes  in a model independent way 
the break-up point. Strangeness is nearly (within error)  balanced. 

In the first column of table \ref{newfitqpbs}
 we see that imposing exact strangeness balance increases
the chemical freeze-out temperature $T$ slightly 
from 145 to 150 MeV. Insisting  on exact balance is 
an improper procedure since the WA97 central rapidity data, which are 
an important input into this analysis, are only known at central rapidity.
It is likely that the longitudinal flow of light quark content contributes to 
some mild $s$--$\bar s$-quark separation in rapidity. For this reason we 
normally consider the results presented in right 
column of table \ref{newfitqpbs} to be more representative of 
the freeze-out dynamics in Pb--Pb 
interactions at central rapidity at $\sqrt{s_{NN}}=17.2$ GeV.

\begin{table}[tb]
\caption{\label{newfitqpbs}
Physical properties of Pb--Pb 158$A$ GeV fireball, left 
column with and right column without imposed strangeness balance. 
We do not include $\Omega+\overline\Omega$ yields in this analysis, 
see end of section \protect\ref{freezespectra}. For more details see text.}
\vspace{-0.2cm}\begin{center}
\begin{tabular}{l|c|c}
                           &Pb$|_v^{\rm s,\gamma_q}$ &Pb$|_v^{\rm \gamma_q}$\\
$\chi^2_{\rm T};\ N;p;r$   &2.25;\ 10;\,3;\,2 &\hspace*{-0.5cm} 1.36;\ 10;\,4;\,2 \\
\hline
$T$ [MeV]             &150 $\pm$ 3     &145 $\pm$ 3.5 \\
$v$                     &0.57 $\pm$ 0.04  &0.52 $\pm$ 0.055\\
$\lambda_{q}$             &1.616 $\pm$ 0.025&1.625 $\pm$ 0.025\\
$\lambda_{s}$             &1.105$^*$         &1.095 $\pm$ 0.02\\
$\gamma_{q}$  & ${\gamma_{q}^c}^*\!=\!e^{m_\pi/2T_f}\!=\!1.61$  &${\gamma_{q}^c}^*\!=\!e^{m_\pi/2T_f}\!=\!1.59$\\
$\gamma_{s}/\gamma_{q}$   &1.02 $\pm$ 0.06  &1.02 $\pm$ 0.06\\
\hline 
$E_{f}^{in}/S_{f}$   &0.163 $\pm$ 0.01    &0.158 $\pm$ 0.01\\
${s}_{f}/b$               &0.68 $\pm$ 0.05  &0.69 $\pm$ 0.05\\
$({\bar s}_f-s_f)/b\ \ $  &0$^*$            &0.05 $\pm$ 0.05\\ 
\end{tabular}
\end{center}
\vskip -0.8cm
\end{table}

\subsection{RHIC freeze-out}\label{freezeRHIC}
There is now first hadronic particle and 
strangeness data from RHIC $\sqrt{s_{NN}}=130$ GeV,
presented at QM2001 by the STAR collaboration \cite{STARkstar}. 
We draw the following conclusions from these results, which in part 
agree with concluding remarks by Nu Xu made at QM2001 \cite{NuX01};
\begin{enumerate}
\item  from $\bar p/p=0.6\pm0.02=\lambda_q^{-6}$ 
it follows $\lambda_q=1.09$;
\item  and hence   $\mu_B=38$ MeV  (18\% of SPS value) at $T=150$ MeV.
If a hadronization at  $T=175$ MeV applies this value rises to $\mu_B=44$ MeV.
\item The ratios
$\overline\Lambda/\Lambda=0.73\pm 0.03=\lambda_s^{-2}\lambda_q^{-4}$ and
$\overline\Xi/\Xi=0.82\pm 0.08=\lambda_s^{-4}\lambda_q^{-2}$ are consistent
within 1.5\% with $\lambda_s=1$,  value expected for sudden hadronization.
\item $K^-/K^+=0.88\pm 0.06$  is also consistent within error with $\lambda_q=1.09$.
\item On the other hand
the ratio  $K^*/\overline{K^*}\simeq 1$ differs from $K/\overline{K}$
significantly. This suggests that $K^*,\overline{K^*} $ yields are
influenced at the level of 10\% 
by `in hadronization' decay product re-scattering 
in an asymmetric way. 
\item Thus
$K^*, \overline{K^*}$ should not be used to fix $T$ 
using the ratios $ K^* /h^-$ and  $ \overline{K^*} /h^-$.
\item The ratio  $ \bar p/\pi=8\%$ cannot be used to fix $T$ since
the $ \bar p$ yield contains undetermined hyperon feed \cite{Raf99}.
\item  The ratio $K^-/\pi^-$ does not suffice to fix the
temperature: we need at least 3 reliable yield ratios as we must also
fix $\gamma_q,\gamma_s$:  $K^-/\pi^-=15\%=f(T)\gamma_s/\gamma_d$.
\end{enumerate}
We find that the first RHIC results allow to understand the magnitude of
chemical potentials ($\mu_s=0, \mu_b=38$ MeV), 
but $T$ and $\gamma_q,\gamma_s$ cannot yet be fixed.
Given the re-scattering phenomena of resonances, 
see section \ref{resonance}, one cannot do a global
analysis without stable strange hadron yields, akin to the situation
we have at the SPS energy range. Thus the final analysis must await the time 
these results become available.  On the other hand the strong 
presence of observable resonances in the hadronic
final state reported by the STAR experiment, 
implies that hadronization has occurred in a sudden fashion, as has 
been seen at SPS. Other  RHIC results, such as particle 
correlation analysis, are
also strongly suggestive of sudden break-up/hadronization.

The most interesting  departure at RHIC from
SPS physics is the great strangeness density. 
We note that:
$$
\frac{dN_{K^+}}{dy}\vert_{y=0}=35\pm3.5\,,
\quad 
\frac{dN_{K^-}}{dy}\vert_{y=0}=30\pm3\,.
$$
Total strangeness $(\bar s)$ 
yield depends on unmeasured hyperons. Model calculations 
suggest more than 20\%. Hence:
$$
\frac{d\bar s}{dy}\vert_{y=0} > 85\pm9\,.
$$
Compare this to: 
$$\frac{d\pi^+}{dy}\simeq  \frac{d\pi^-}{dy} \simeq 235.$$ 
Under these conditions calculations suggest that 
$\bar s/b\simeq 8$ (11--12 times greater compared to
 17$A$ GeV SPS Pb--Pb). 

Given this  strangeness rapidity yield it is very 
difficult to imagine that among three (anti)quarks which coalesce to make
a (anti)baryon there is  no (anti)strange quark. Hence we found 
in a statistical model study that most
baryons and antibaryons produced will carry strangeness \cite{Raf99}. 
Thus currently observed 
non-strange nucleons and antinucleons are strongly contaminated 
by hyperon decay feed, and at this time the reported nucleon RHIC results
cannot be used in order to characterize freeze-out conditions. 
Corrections as large as factor 2--3 in relative yields must be expected.
The influence of this effect on {\it e.g.} antiproton-$m_\bot$ spectra has so 
far not been  quantitatively explored.

\subsection{Strange hyperon $m_\bot$ spectra}\label{freezespectra}
About a year ago the experiment WA97 
determined the relative normalization of $m_\bot$-distribution for
strange particles $\Lambda,\,\overline\Lambda, \,\Xi,\,
\overline\Xi,\, \Omega+ \overline\Omega, \, K_s=(K^0+\overline{K^0})/2$
in four  centrality bins \cite{Ant00}.
We have since obtained a simultaneous description of the absolute yield 
(chemical freeze-out) and shape (thermal freeze-out) of these results \cite{Tor00}. 
Our strategy has been to maximize the precision of the 
description of the final multi-particle hadron state employing
statistical methods. 

This requires that we introduce parameters which characterize
possible chemical non-equilibria, and  velocities  of matter
evolution. Our Analysis employed two velocities: a local  
flow velocity $v$ of the fireball 
volume element where from particles emerge,
and hadronization surface (breakup) velocity which we refer to 
as $v_f^{\,-1}\equiv dt_f/dx_f$.

We have found, as is generally believed and expected, 
 that all hadron $m_\bot$-spectra are strongly influenced by 
resonance decays. In the spectral analysis we assume 
 that decay products of resonances
do not  reequilibrate in rescattering, i.e. there is a tacit 
assumption that the freeze-out is sudden, and thus we can only
test for consistency of this approach. The final
particle distribution is composed of directly produced particles
and non-rescattered first generation decay products,
as no other contributing decays are known for hyperons, and 
hard kaons. 

Since resonance contributions are important, the
correct combination of the direct and decay contributions 
influences the  detailed shape of the spectra, and thus one can 
determine the freeze-out temperature alone from the study of 
the single particle $m_\bot$ spectra, once these are very precisely 
known. This approach fixes a best temperature and velocity
of expansion and hadronization without any additional input,
such as is HBT correlation analysis, commonly used in this context.

Our procedure to determine the combined spectrum 
was based on Ref.  \cite{Sch95,Ani85}.
The best statistical parameters which minimize the total relative
error $\chi^2_{\rm T}$ at 
a given centrality is than determined fitting all available
spectral data points keeping the different collision 
centrality apart.

The results of the thermal analysis are in excellent agreement
with the chemical analysis. In all centrality bins we find 
that the thermal freeze-out temperature $T$ is in agreement
with the chemical freeze-out condition. 
There is  no indication 
of a significant or systematic change of $T$ with centrality. 
This is consistent with the believe that the formation of the new state of 
matter at CERN is occurring in all centrality bins explored by the 
experiment WA97. 
It will be interesting to see if the low centrality 5th bin now
studied by experiment NA57 will show different 
characteristics \cite{NA57Mor}.

The magnitude of the 
collective expansion velocity $v$  is also found
to be in excellent agreement with the 
chemical freeze-out analysis. 
Though within the experimental error,  there is found a
systematic increase in transverse flow velocity $v$ with centrality and thus 
size of the system. This is expected, since the more central events comprise 
greater volume of matter, which allows more time for development
of the flow.

The chemical analysis has not been
sensitive to the  break-up (hadronization) speed 
parameter $v_f$,   which was for the first time 
determined in the thermal analysis. 
The value of $v_f$ found is near to 
velocity of light which is   
consistent with the picture of a 
sudden breakup of the  fireball.

The strange  particle $m_\bot$-spectra  of 
$\Lambda,\,\overline\Lambda, \,\Xi,\,
\overline\Xi,\, K_s=(K^0+\overline{K^0})/2$ ) are
reproduced in great precision and without 
systematic variations, but $\Omega+\overline\Omega$.
Although in the purely chemical fit discussed in
subsection \ref{freezechem} we excluded the $\Omega,\overline\Omega$
yields due to their anomalous production pattern, we did include their spectra
in the thermal analysis.  In  all four centrality bins for the
 sum $\Omega+\overline\Omega$ we systematically  under predict the two lowest
$m_\bot$ data points. This low-$m_\bot$ excess also explains why the 
inverse $m_\bot$ slopes for $\Omega,\overline\Omega$ 
are reported to be smaller than the values seen in all other strange (anti)hyperons.

The 1.5 s.d. low $p_\bot$ deviations when summed over all
bins of the $\Omega+\overline\Omega$ 
spectrum translates into 3 s.d. 
deviations from the prediction of the statistical model chemical analysis. 
It has been proposed that this excess is evidence, but not proof,
that $\Omega,\overline\Omega$  are produced as topological defects 
arising from the formation of disoriented chiral condensates (DCC) 
with an average domain size of about 2 fm \cite{Kap00}. However, an
excess above statistical yield is also expected due to in source
(anti)strange quark correlations \cite{Raf82},  visible 
in the hadron of smallest statistical yield, 
such as is  $\Omega,\overline\Omega$. For further details of the
thermal fit the reader should consult Ref.
\cite{Tor00}.

\section{Resonances and freeze-out dynamics}\label{resonance}
We explore here if it is possible to  experimentally determine 
the period of time a fireball particle 
is in touch with matter  after formation 
and before it is free-streaming, 
using strange hadron resonances \cite{Tor01}.
At this time $\Lambda(1520)$ of 
width $\Gamma_{\Lambda(1520)}=15.6$ MeV
has been observed in heavy ion reactions at SPS energies 
 \cite{NA49Res,NA49Res2}.
Both SPS \cite{NA49Res2} and RHIC experiments \cite{STARkstar}
report measurement of the $K^*(892)$ signal, which has a
much greater width, $\Gamma_{K^*(892)}=50$ MeV. 

The $\Lambda(1520)$ abundance yield is found about 2 times smaller
than expectations based on the yield 
extrapolated from nucleon-nucleon reactions, scaled with hadron yield. 
This is  to be compared with an increased production by factor 2.5 of 
$\Lambda$.  A possible explanation for
this relative suppression by a factor 5 is that the
decay products ($\pi,\Lambda$)  have re-scattered and thus their momenta
did not allow  to reconstruct this state in an invariant mass analysis.
However, the observation of a strong $K^*$-yield signal  
contradicts this point  of view, since this is a faster decaying resonance:
a back of envelope calculation based 
on exponential population attenuation and assuming that all
decays in matter become unobservable
suggests that if the observable yield of  $\Lambda(1520)$ is reduced 
by factor 5, the observable 
yield of $K^*(892)$  should be suppressed by a factor 15.
This is clearly not the case, as the $K^*(892)$ yield is significant.

Another explanation is that in matter $\Lambda(1520)$
decays faster and there is much more opportunity for the rescattering of decay 
products, and  fewer observable resonances.
The width of $\Lambda(1520)$ can be quenched in collisions such as
$$ \pi+\Lambda(1520)\to \Sigma^*\to   \pi+\Lambda\,,$$
since $\Gamma_{\Lambda(1520)}$  is  small due to
need for angular $L=2$ partial wave in its decay.  Collisional widening of 
a meta stable state is a familiar phenomenon explored in several 
areas of physics \cite{quench}. The decay of the $\phi(s\bar s)$ 
has been the `usual suspect' in search for such a quenching, 
given the proximity of the $K\overline K$ 
mass threshold \cite{Bi91,Ko92,Kli98}. It should be noted that the 
experimentally observed width will always be the natural width, since 
in-matter-decay products are not allowing  $\Lambda(1520)$ reconstruction
(see below).

The observable yield of resonances is thus controlled by several physical
properties, such as the freeze-out temperature $T$, the decay width in 
matter $\Gamma$, and  the time spend in the hadron phase after freeze-out $\tau$. 
The suppressed yield  can 
mean either a low temperature chemical freeze-out, or a long interacting
phase with substantial re-scattering. 
We have formulated a simple model based on the width of the resonances in question 
and the decay products reaction cross-sections  within an
expanding fireball of nuclear matter. 
For more details  we refer to
Ref.\,\cite{Tor01}. 

We found that the observable resonance yields
are very sensitive to the interaction period  in the hadron phase,
but not to magnitude of interaction cross sections used. It turns out that
practically all resonances which decay inside matter become unobservable,
the medium is opaque as it scatters effectively even at realtaively small
cross sections the decay products. The observable resonance yield can 
be  derived from original $T$-controlled yield followed by a  
comparison of the lifespan of the hadron phase and decay lifespan of the 
resonance in medium. 
 
For $\Lambda(1520)/(\mbox{all }\Lambda)$ 
we show the result in Fig.\,\ref{LamRes}:  the bottom
portion is for the natural width $\Gamma_{\Lambda(1520)}=15.6$ MeV,
and the top portion of the figure is for a width quenched to 
$150$ MeV. We recall that in the just completed study of NA49 experiment 
\cite{NA49Res} has found $\Lambda(1520)/(\mbox{all }\Lambda)=0.025\pm 0.008$ 
which is barely if at all compatible with the unquenched result, since it implies
an extremely long hadronization time of about $20\pm5$ fm/c (depending on
freeze-out temperature) which is incompatible with other 
experimental results. On the other hand,  we see in the top
portion of Fig.\,\ref{LamRes} that after introduction of a quenched 
resonance width  the experimental result 
is compatible for all freeze-out temperatures with 
a sudden hadronization model -- the magnitude of the freeze-out time 
(1 fm/c) is a consequence
of the assumed  quenched width, suggested by the phase space size of the 
decay  once angular momentum selection rule in the 
decay is overcome. This value of the width can be a factor two different without
altering the physical conclusion.

This finding tells us that  a study  of
several resonances with considerably different physical
properties must be  used in an investigation of freeze-out dynamics of QGP.
Among strange particle resonances, 
the  $\Sigma^*(1385)$, with $\Gamma_{\Sigma^*(1385)}=35$ MeV is 
in our opinion most interesting. This state which decays primarily
into $\Lambda$ is on theoretical grounds produced
an order of magnitude more abundantly than is $\Lambda(1520)$, due to a high 
degeneracy factor and smaller mass. 
Without in medium quenching,  the $\Sigma^*$ signal 
is  more strongly influenced by final state interactions than that
of  $\Lambda(1520)$, but not as strong as $K^*(892)$.  

We can express our finding representing
one resonance yield (normalized ratio)
 against the other.  as is seen in  
Fig.~\ref{projSigK}.
As indicated from top to bottom in the grid, the 
lifespan in fireball increases, while from
left to right the temperature of chemical particle freeze-out increases. 
The medium is effectively opaque, all resonances that decay
in medium become unobservable. 
A remarkable result is found  for unquenched resonances 
$\Sigma^{*}$/(all $\Lambda$) with  
$K^*(892)/$(all K), seen in bottom of  Fig.\,\ref{projSigK}.
This projection results in a nearly 
unique line in the two dimensional plane, and thus 
any deviation from this result constitutes a firm 
evidence for resonance quenching. This is seen  in the 
top portion of Fig.\,\ref{projSigK} obtained with 
a quenched $\Gamma_{\Sigma^{*}}=150$.

\begin{figure}[tb]
\begin{center}
\hspace*{.1cm}
\psfig{width=8cm,clip=1,figure=\pathnow 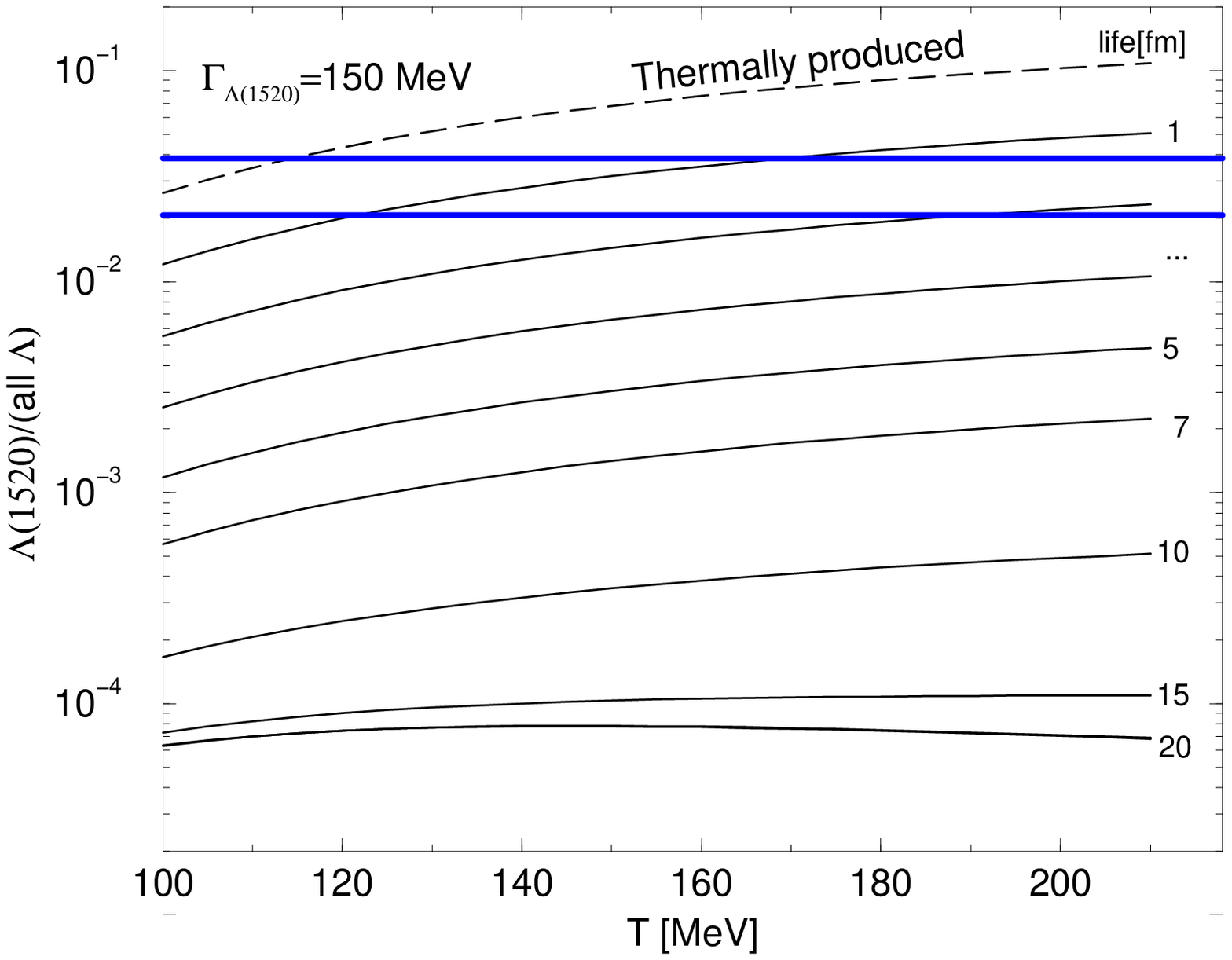}\\
\vspace*{0.1cm}
\psfig{width=8cm,clip=1,figure=\pathnow 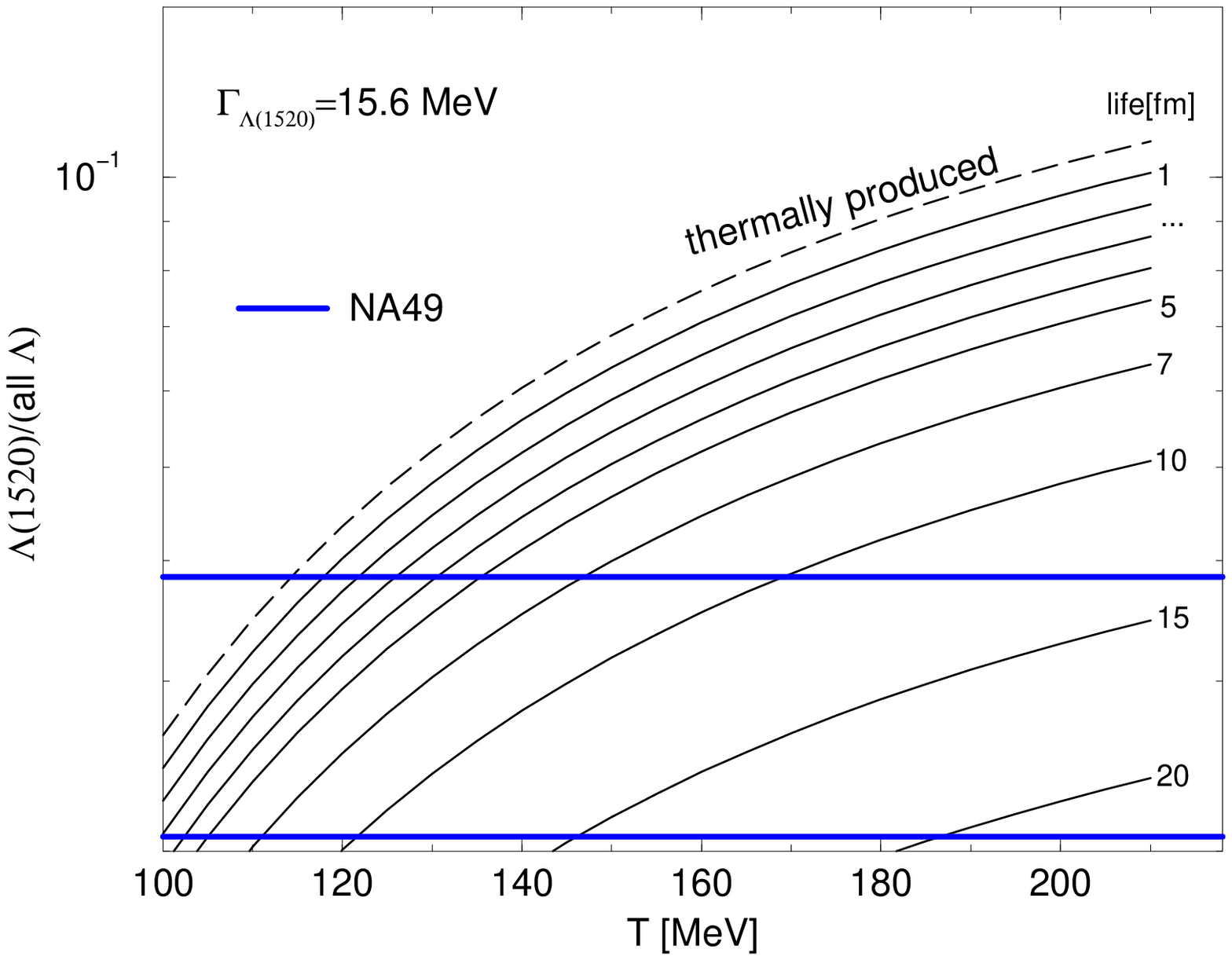}\\
\vspace*{0.1cm}
\caption{Relative $\Lambda(1520)/(\mbox{all }\Lambda)$ yield as function
of freeze-out temperature $T$. Dashed - thermal yield, solid lines: 
observable yield for evolution lasting the time shown (1....20 fm)
in an opaque medium. Horizontal lines: experimental limits of
NA49 \protect\cite{NA49Res}.  Bottom: natural resonance width 
$\Gamma_{\Lambda(1520)}=15.6$ MeV, 
top: quenched $\Gamma_{\Lambda(1520)}=150$ MeV.  
\label{LamRes}}
\end{center}
\end{figure}
\vskip -0.3cm

\section{Summary and conclusions}
We updated our chemical freeze-out analysis and have
compared with the hyperon thermal freeze-out analysis  in 
subsection \ref{freeze}. Our results  confirm that
CERN-SPS  results originate in interesting and new 
physics phenomenon, and is consistent with the reaction picture of
a suddenly hadronizing QGP-fireball \cite{Raf00}, since both
chemical and thermal freeze-out coincide. We were able to
determine the thermal freeze-out alone from  a single particle spectra since
the spectrum includes heavier resonance contribution.
A similar analysis of the $m_\bot$ spectra for high energy collisions 
has been carried out recently \cite{Bec00}. This work 
reaches for elementary high energy processes similar conclusions 
as we have presented regarding the identity of 
chemical and thermal freeze-out. The higher  freeze-out temperature 
found in elementary interactions 
is also consistent with our results, considering that 
only in nuclear collisions significant 
super cooling is expected \cite{Raf00}. 
In our view the large nuclear collision (quark-gluon?) fireball is driven
to rapid expansion by internal pressure, and 
ultimately  a sudden breakup (hadronization) 
into final state  particles occurs which reach detectors 
without much, if any, further re-scattering. The required
sudden fireball breakup  arises since as the fireball super-cools, and in this 
state encounters a strong mechanical instability \cite{Raf00}. Note that
deep super cooling requires a  first order phase transition.

\begin{figure}[tb]
\begin{center}
\hspace*{0.2cm}
\psfig{width=8cm,clip=1,figure=\pathnow 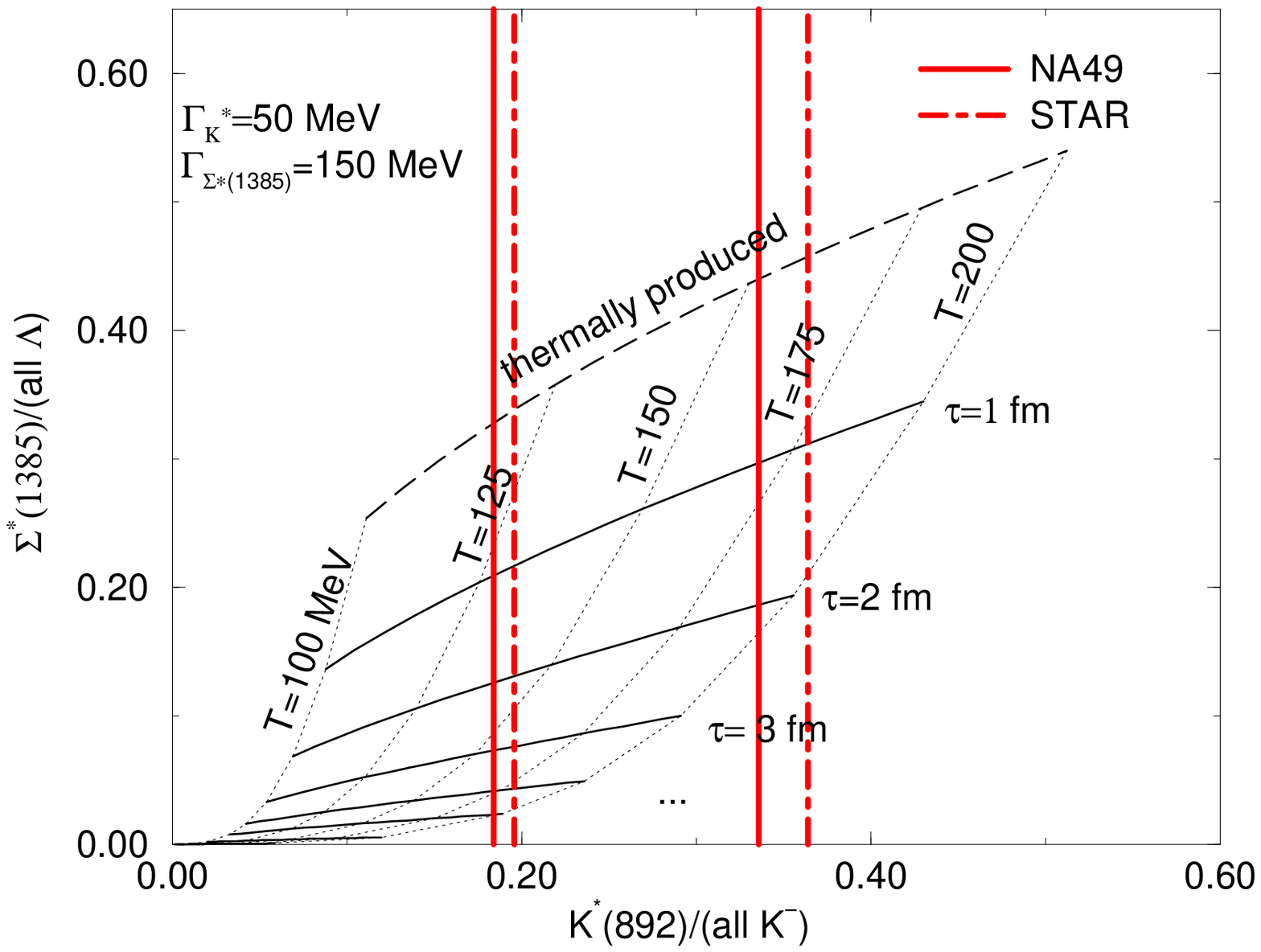}
\vspace*{0.21cm}
\hspace*{-0.cm}
\psfig{width=8cm,clip=1,figure=\pathnow 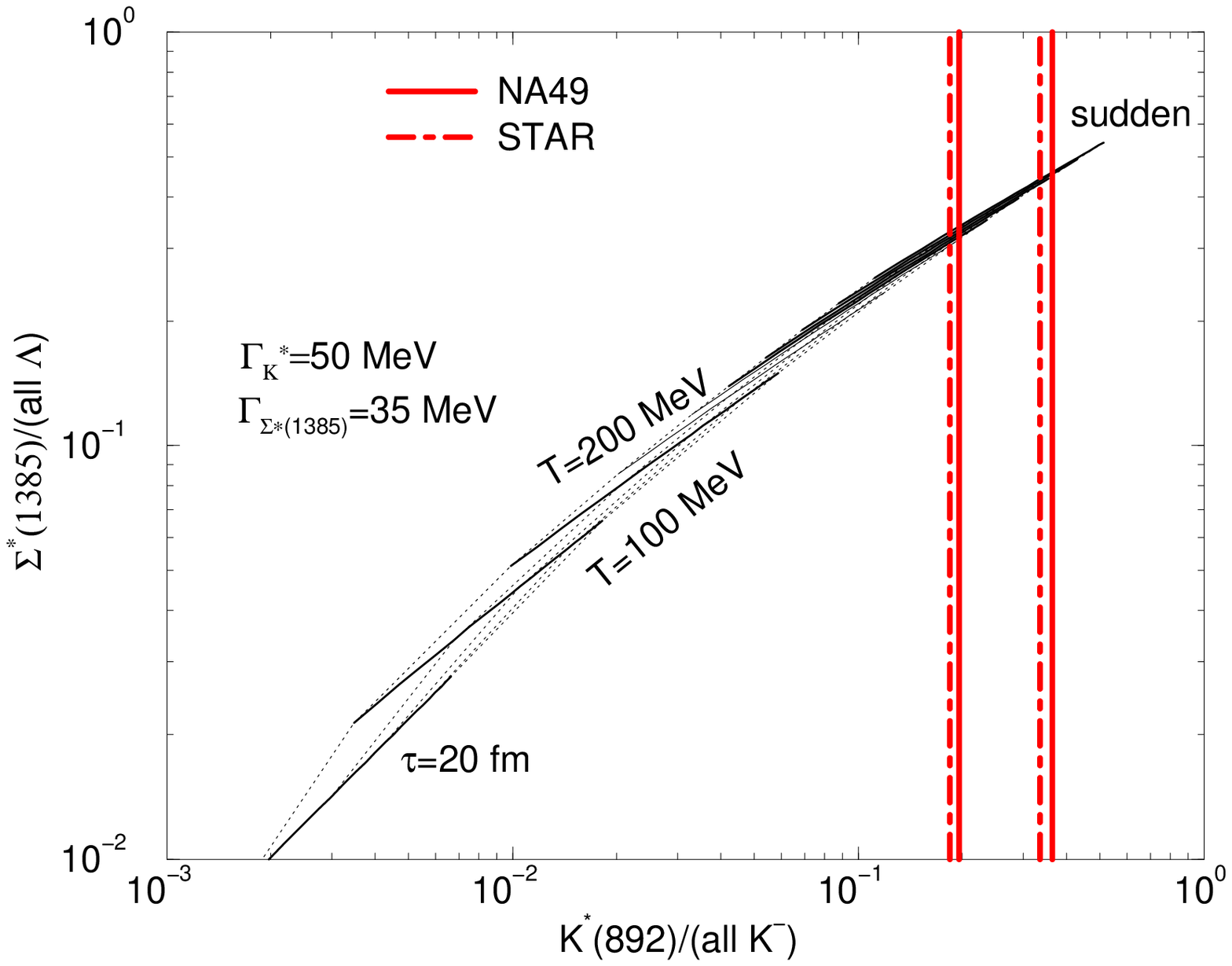}
\caption{Dependence of the combined 
$\Sigma^{*}$/(all $\Lambda$) with  
$K^*(892)/$(all K) 
signals on the chemical freeze-out temperature
and interacting phase lifetime. Top: quenched $\Gamma_{\Sigma^{*}}=150$
MeV, bottom natural widths. Vertical lines: experimental limits of
NA49  \protect\cite{NA49Res}.}
\label{projSigK}
\end{center}
\end{figure}

We have presented in section \ref{resonance} results 
on strange hadron resonance production 
which allow to study the dynamics of thermal and chemical freeze-out.
A comparison of
several resonances with considerably different physical
properties must be  used in a study of freeze-out dynamics of QGP.
Strange resonances are easier to explore, since their decay 
involve rarer strange hadrons and thus the backgrounds are smaller. 
Moreover, the detectability of the naturally wide non-strange
resonances is always relatively small, except if (very) sudden hadronization
applies. For this reason it will be quite interesting
to see if  $\Delta(1230)$ can be observed at all, as this
would be only possible if  chemical and thermal freeze-out 
conditions are truly coincident. 

The observability of several  strange hadron resonances depends if these
decay in matter or outside. The more short lived a resonance is, the
more likely it is to decay within the confined hadron matter period
of fireball evolution. Suitably comparing yields of several 
resonances we can hope to resolve the question how
sudden hadronization of QGP in fact is. 
We studied the suppression of observability of three strange resonances  
$\Lambda(1520), K^*(892), \Sigma^*(1385)$   as a tool capable
of estimating conditions at particle freeze-out.  
Our objective was to quantify  how temperature, lifespan and 
the (quenched) width $T,\tau,\Gamma_i$  for the 
resonance $i$ influence the observable yield.  $\Gamma_i$ 
in matter may significantly differ from natural width.

This discussion of how resonances help to 
understand the hadronization dynamics is a beginning of a complex
analysis which will occur in interaction with experimental 
results. We saw  that observable strange resonance yields can
vary widely depending on conditions which should allow a detailed study of 
QGP freeze-out dynamics. We believe considering $\Lambda(1520)$ result 
that in-matter resonance lifetime quenching is significant.

\vskip 0.3cm
\noindent{\bf Acknowledgments:}
We thank Federico Antinori, Christina Markert, Emanuele Quercigh, 
Chrstelle Roy,  Karel Safarik, and Zhangbu Xu,  for valuable comments.
Work supported in part by a grant from the U.S. Department of
Energy,  DE-FG03-95ER40937. Laboratoire de Physique Th\'eorique 
et Hautes Energies, University Paris 6 and 7, is supported 
by CNRS as Unit\'e Mixte de Recherche, UMR7589.


\end{narrowtext}
\end{document}